\documentclass[10pt, prb, aps, twocolumn, showpacs, citeautoscript, floatfix, reprint, amsmath, amssymb, notitlepage, superscriptaddress, longbibliography]{revtex4-2}

\usepackage{graphicx}
\usepackage{stmaryrd}
\usepackage[countmax]{subfloat}
\usepackage{comment}
\usepackage{physics}
\usepackage[hidelinks]{hyperref}
\usepackage{rotating}
\usepackage{amsmath}
\usepackage{amsfonts}
\usepackage{amssymb}
\usepackage{wasysym}
\usepackage{dcolumn} 
\usepackage{bm} 
\usepackage{color}

\usepackage{float}
\usepackage{latexsym}

\begin{document}

\title{Topological marker in three dimensions based on kernel polynomial method}

\author{Ranadeep Roy}
\affiliation{Department of Physics, The Ohio State University, Columbus OH 43210, USA}

\author{Wei Chen}
\affiliation{Department of Physics, PUC-Rio, 22451-900 Rio de Janeiro, Brazil}

\date{\rm\today}

\begin{abstract}

The atomic-scale influence of disorder on the topological order can be quantified by a universal topological marker, although the practical calculation of the marker becomes numerically very costly in higher dimensions. We propose that for any symmetry class in higher dimensions, the topological marker can be calculated in a very efficient way by adopting the kernel polynomial method. Using class AII in three dimensions as an example, which is relevant to realistic topological insulators like Bi$_{2}$Se$_{3}$ and Bi$_{2}$Te$_{3}$, this method reveals the criteria for the invariance of topological order in the presence of disorder, as well as the possibility of a smooth cross over between two topological phases caused by disorder. In addition, the significantly enlarged system size in the numerical calculation implies that this method is capable of capturing the quantum criticality much closer to topological phase transitions, as demonstrated by a nonlocal topological marker.

\end{abstract}

\maketitle

\section{Introduction} 

The influence of disorder on the topological order has been an important issue that are detrimental to various experimental measurements and is oftentimes unavoidable. Naively, the effect of disorder is expected to depend on the spatial dimension and symmetry of the material. For instance, the local response to impurities is usually more dramatic in lower dimensions\cite{Anderson58,Abrahams79,Lee85}, and symmetry-breaking impurities may have a more severe influence than symmetry-preserving ones, e.g., the backscattering caused by magnetic impurities in time-reversal symmetric topological materials\cite{Qi08,Liu09,Jack20}. To address these issues, the method of topological marker has been proposed\cite{Bianco11,Prodan10,Prodan10_2,Prodan11}, which maps the topological invariant defined in momentum space to individual lattice sites in real space, enabling  one to address how the topological order is modified locally around impurities. Furthermore, based on a universal topological invariant for Dirac models in momentum space\cite{vonGersdorff21_unification}, a universal topological marker has been proposed\cite{Chen23_universal_marker}, which allows to investigate topological insulators (TIs) and topological superconductors (TSCs) in any dimension and symmetry class\cite{Schnyder08,Ryu10,Kitaev09,Chiu16}. However, despite posing as a ubiquitous tool to investigate the influence of disorder down to atomic scale, the practical calculation of the universal topological marker is numerically very costly in higher dimensions, especially if one aims to simulate a large lattice in order to capture the fine details around individual impurities, which hinders the investigation of disorder effect in higher dimensions.

To overcome the cost of numerical calculation, the kernel polynomial method (KPM) emerges as a numerically efficient method \cite{Weisse06}. The feasibility of this method applied to topological materials has been demonstrated in the calculation of quantized Hall conductivity in disordered Chern insulators\cite{Garcia15}. Within the context of the topological marker formalism, the KPM has been applied to the three-dimensional (3D) topological crystalline insulators characterized by a mirror Chern number\cite{Varjas20}, the Chern marker in the two-dimensional (2D) disordered Haldane model\cite{Romeral25}, as well as the second and third Chern number in four- and six-dimensional Chern insulators\cite{Chen25}, in which the KPM approximates the projectors onto lattice eigenstates in the calculation of the markers by a polynomial expansion. Motivated by these pioneering works, in this paper we propose that the KPM can also be applied to the projectors in the calculation of the aforementioned universal topological marker to reduce the numerical effort, thereby significantly broadens the application of the KPM to any dimension and symmetry class. Besides, we demonstrate that the KPM also captures the off-diagonal elements of the marker that has been called the nonlocal topological marker, since it depends on the relative position between two lattice sites\cite{Molignini23_Chern_marker,Chen23_spin_Chern,Chen23_universal_marker}. The significance of this nonlocal marker is that it serves as an appropriate correlation function to quantify the critical behavior near topological phase transitions, with a decay length that diverges at the critical point\cite{Chen17,Chen19_universality}. Because the system size must be larger than the correlation length in order to capture the decay of the nonlocal marker, the fact that KPM enables the numerical calculation at a significantly larger system size means that the critical behavior very close to the critical point can be captured by this method, pointing to another advantage of KPM in addressing the topological quantum criticality.

As a concrete example, we use class AII in three dimensions (3D) to demonstrate our formalism, which is a particularly important symmetry class that describes realistic TIs like Bi$_{2}$Se$_{3}$ and Bi$_{2}$Te$_{3}$\cite{Zhang09,Liu10}. Armed with this numerically efficient tool, we uncover that if the impurities correspond to altering the nonzero matrix elements of the lattice Hamiltonian, regardless of whether the symmetry is broken or not, the spatially averaged marker remains unchanged at small impurity density, signifying the robustness of topological order against this type of disorder. In addition, we find that impurities can also be used to smoothly change the topological invariant from one integer to another, in contrast to the abrupt jump of the topological invariant in homogeneous systems caused by band inversion\cite{Hasan10,Qi11}. These features are very similar to those in one and two dimensions uncovered recently\cite{Oliveira24}, suggesting that all TIs and TSCs share very similar responses to disorder regardless of the dimension and symmetry class. Finally, by combining the KPM with an exponentiated position operator, we present the smoothly decaying profile of the nonlocal topological marker, and the evidence of a diverging decay length as the system gradually moves towards the critical point, thereby demonstrating the advantage of this method in quantifying the topological quantum criticality.

\section{Universal topological marker in higher dimensions}

\subsection{Kernel polynomial method applied to universal topological marker}

Our aim is to elaborate the numerical efficiency of the KPM in the calculation of topological marker in higher dimensions. For this purpose, we first review the formalism of the universal topological marker applicable to Dirac models in any dimension and symmetry class. For a $D$-dimension TI or TSC described by the Dirac Hamiltonian in momentum space, the Hamiltonian $H_{0}({\bf k})={\bf d}({\bf k})\cdot{\boldsymbol\Gamma}$ is expanded by the Dirac matrices $\Gamma_{i}=(\Gamma_{0},\Gamma_{1}...\Gamma_{2n})$ of dimension $2^{n}\times 2^{n}$ that satisfy the Clifford algebra  $\left\{\Gamma_{i},\Gamma_{j}\right\}=2\delta_{ij}$, and ${\bf d}({\bf k})=(d_{0},d_{1}...d_{D})$ describes the momentum dependence. The corresponding real space lattice Hamiltonian $H_{0}$ is then obtained from a Fourier transform of $H_{0}({\bf k})$, which in the second quantized form contains the electron operators $c_{i}$ with $i=\left\{{\bf r},\nu\right\}$, where ${\bf r}$ is the position of the unit cell, and $\nu$ denotes the degrees of freedom inside a unit cell such as spin and orbital. The universal topological operator constructed out of $H_{0}$ takes the form of two arrays of alternating projectors $\left\{P,Q\right\}$ sandwiched by position operators ${\hat r}_{1\sim D}=\left\{{\hat x},{\hat y},{\hat z}...\right\}$
\begin{eqnarray}
{\hat {\cal C}}=N_{D}W\left[Q\,{\hat r_{1}}P\,{\hat r_{2}}...\,{\hat r_{D}}{\cal O}+(-1)^{D+1}P\,{\hat r_{1}}Q\,{\hat r_{2}}...{\hat r_{D}}{\overline{\cal O}}\right],
\nonumber \\
\label{topological_operator}
\end{eqnarray}
where $W=\Gamma_{D+1}\Gamma_{D+2}...\Gamma_{2n}$ is the product of all the unused Dirac matrices expressed in the lattice basis, and the last operators are given by $\left\{{\cal O},\overline{\cal O}\right\}=\left\{P,Q\right\}$ if $D$ is odd, or $\left\{{\cal O},\overline{\cal O}\right\}=\left\{Q,P\right\}$ if $D$ is even due to the alternating order of $P=\sum_{n}|E_{n}\rangle\langle E_{n}|$ and $Q=1-P=\sum_{m}|E_{m}\rangle\langle E_{m}|$ that are projectors to the filled $|E_{n}\rangle$ and empty $|E_{m}\rangle$ lattice eigenstates\cite{Chen23_universal_marker}. The normalization factor is $N_{D}=i^{D}2^{2D-n}\pi^{D}/c\,V_{D}$, with $V_{D}=\left\{V_{1},V_{2},V_{3}...\right\}=\left\{2\pi,4\pi,2\pi^{2}...\right\}$ the volume of a $D$-sphere, and the value of the prefactor $c={\rm Tr}\left[\Gamma_{0}\Gamma_{1}...\Gamma_{2n}\right]/2^{n}=\left\{1,-1,i,-i\right\}$ depends on the representation of $\Gamma$-matrices for the specific TI or TSC under question. To be more specific, the topological operator from 1D to 4D is 
\begin{eqnarray}
&&{\hat{\cal C}}_{1D}=N_{1D}W\left[Q{\hat x}P+P{\hat x}Q\right],
\nonumber \\
&&{\hat{\cal C}}_{2D}=N_{2D}W\left[Q{\hat x}P{\hat y}Q-P{\hat x}Q{\hat y}P\right],
\nonumber \\
&&{\hat{\cal C}}_{3D}=N_{3D}W
\left[Q\hat{x}P\hat{y}Q\hat{z}P+P\hat{x}Q\hat{y}P\hat{z}Q\right],
\nonumber \\
&&{\hat{\cal C}}_{4D}=N_{4D}W
\left[Q\hat{x}P\hat{y}Q\hat{z}P\hat{w}Q-P\hat{x}Q\hat{y}P\hat{z}Q\hat{w}P\right],
\label{C_1D_to_4D}
\end{eqnarray}
and the pattern repeats in higher dimensions. Once the topological operator has been calculated for a specific system, we extract the local and nonlocal topological markers from the diagonal and off-diagonal elements of the operator, respectively, yielding
\begin{eqnarray}
    &&{\cal C}({\bf r})=\sum_{\nu}\langle{\bf r},\nu|{\hat{\cal C}}|{\bf r},\nu\rangle,
    \nonumber \\
    &&{\cal C}({\bf r,r+R})=\sum_{\nu}\langle{\bf r},\nu|{\hat{\cal C}}|{\bf r+R},\nu\rangle.
    \label{local_nonlocal_markers}
\end{eqnarray}
In homogeneous systems, the local marker ${\cal C}({\bf r})$ is expected to recover the integer-valued topological invariant in the thermodynamic limit $1/L^{3}\rightarrow 0$.

Evaluating the local and nonlocal markers requires diagonalization of the lattice Hamiltonian, which becomes prohibitively memory-intensive as system dimensionality increases. However, its simple structure allows efficient evaluation using the Chebyshev polynomial approximation, also known as the KPM\cite{Varjas20,Romeral25,Chen25}, which allows one to use the sparsity of tight-binding Hamiltonians to reduce the memory requirements. In particular, for a system with $M$ orbitals and using a KPM expansion of order $N$, the memory requirement for computing the invariant at a single energy scales only as $M$ (however, the time complexity is still large as it scales as $MN$) \cite{Varjas20}. The basic idea is as follows : since Chebyshev polynomials form a complete set for functions taking values in $[-1,1]$, one first  rescales the Hamiltonian so that its spectrum lies between $[-1,1]$. Then any function $f$ of the Hamiltonian $H$, which depends on some parameter $\lambda$ can be replaced by its KPM expansion up to order $N$ and its action on an arbitrary column vector $\ket{v}$ can be evaluated:
\begin{equation}
    \hat{f}(\lambda,H)\ket{v} = \hat{f}_{\text{KPM}}(\lambda, H)\ket{v}
\end{equation}
\begin{equation}
    \hat{f}_{\text{KPM}}(\lambda, H)\ket{v} = \sum_{n=0}^N g^N_n\mu_n(\lambda)T_n(H) \ket{v}
\end{equation}
Here $T_n$ denotes the $n^{th}$ order Chebyshev polynomial and $g^N_n\mu_n$ represents the modified moment. Due to the recurrence relation satisfied by Chebyshev polynomials, one does not need to evaluate higher powers of $H$ to obtain the above series :
\begin{align*}
    \ket{v_0}     &= \ket{v} \\
    \ket{v_1}     &= H\ket{v_0} \\
    \ket{v_{m+1}} &= 2H\ket{v_m} - \ket{v_{m-1}}
\end{align*}
For the calculation of the local marker, we need to write a KPM expansion for the spectral projector. At $T=0$, the projector is simply a step function:
\begin{equation}
     P = \theta(\epsilon- H)
\end{equation}
Therefore, its expansion up to order $M$ is :
\begin{equation}
    P(\epsilon,H) = \sum_{m=0}^M g_m\mu_m(\epsilon)T_m(H)
\end{equation}
The moments can be calculated explicitly :
\begin{equation}
        \mu_m(\lambda) =  \frac{2}{\pi} \bigg(\frac{1}{1+ \delta_{m,0}}\bigg) \int_{-1}^1 \frac{\hat{f}(\lambda,E)T_m(E)}{\sqrt{1-E^2}}dE
\end{equation}
and we get :
\begin{equation}
\mu_m(\epsilon) = \begin{cases}
    1 - \frac{\cos^{-1}(\epsilon)}{\pi} \, ; \, m=0 \\
    -\frac{2 \sin{(\cos^{-1}(\epsilon)})}{m\pi} \, ; \, m \neq 0
\end{cases}
\end{equation}
Performing a similar expansion for the complementary projector $Q=1-P$, we obtain the moments :
\begin{equation}
\mu_m(\epsilon) = \begin{cases}
    1 - \frac{\cos^{-1}(\epsilon)}{\pi} \, ; \, m=0 \\
    \frac{2 \sin{(\cos^{-1}(\epsilon)})}{m\pi} \, ; \, m \neq 0
\end{cases}    
\end{equation}
These operators can then be applied to the topological operators in Eqs.~(\ref{topological_operator}) and (\ref{C_1D_to_4D}) according to the dimension and symmetry of the system at hand, and the local and nonlocal markers can be subsequently obtained from Eq.~(\ref{local_nonlocal_markers}). 

\begin{figure}[htbp]
    \centering
    \begin{minipage}{0.49\linewidth}
        \centering
        \includegraphics[width=\linewidth,trim=10 0 25 0, clip]{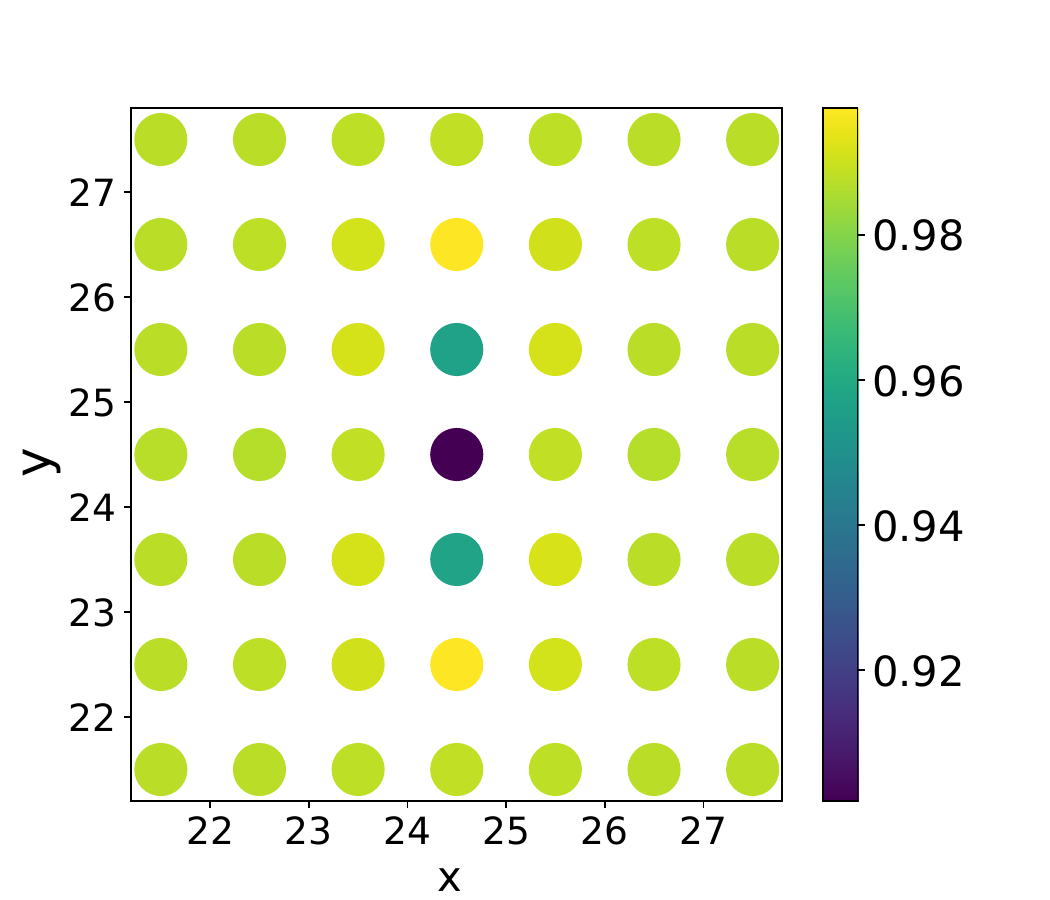}
        \\(a)
    \end{minipage}
    \begin{minipage}{0.49\linewidth}
        \centering
        \includegraphics[width=\linewidth,trim=10 0 25 0, clip]{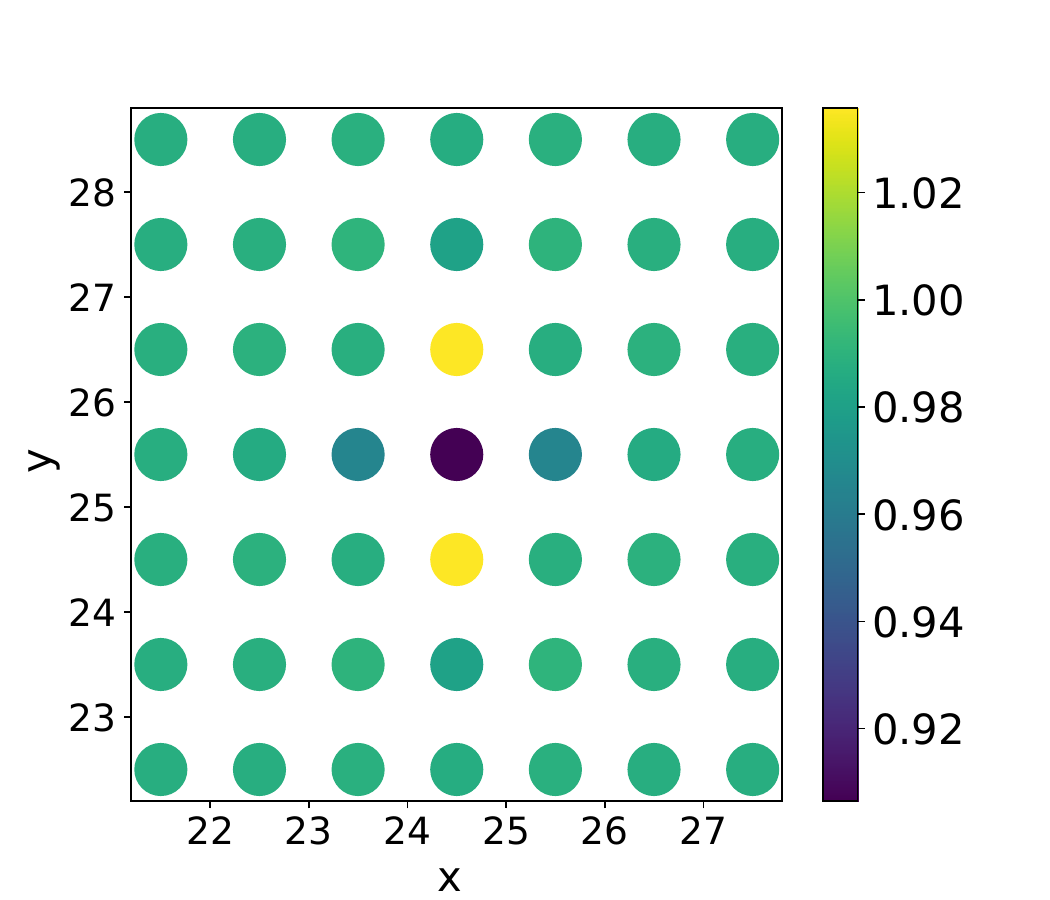}
        \\(b)
    \end{minipage}
    \caption{ Spatial pattern of the marker near the impurity site. The color bar indicates the magnitude of the local marker.In (a), impurity is introduced by modifying the hopping strength ($A_0$) for the hopping from the central site to the nearest neighbour in the $+y$ direction whereas in (b), a single additional hopping to the fourth nearest neighbour (2a units) in $+y$ direction from the central site is introduced.}
    \label{fig:spatial_pattern_marker}
\end{figure}

\begin{figure*}[t]
    \centering
    \begin{minipage}[b]{0.32\textwidth}
        \centering
        \includegraphics[width=\linewidth]{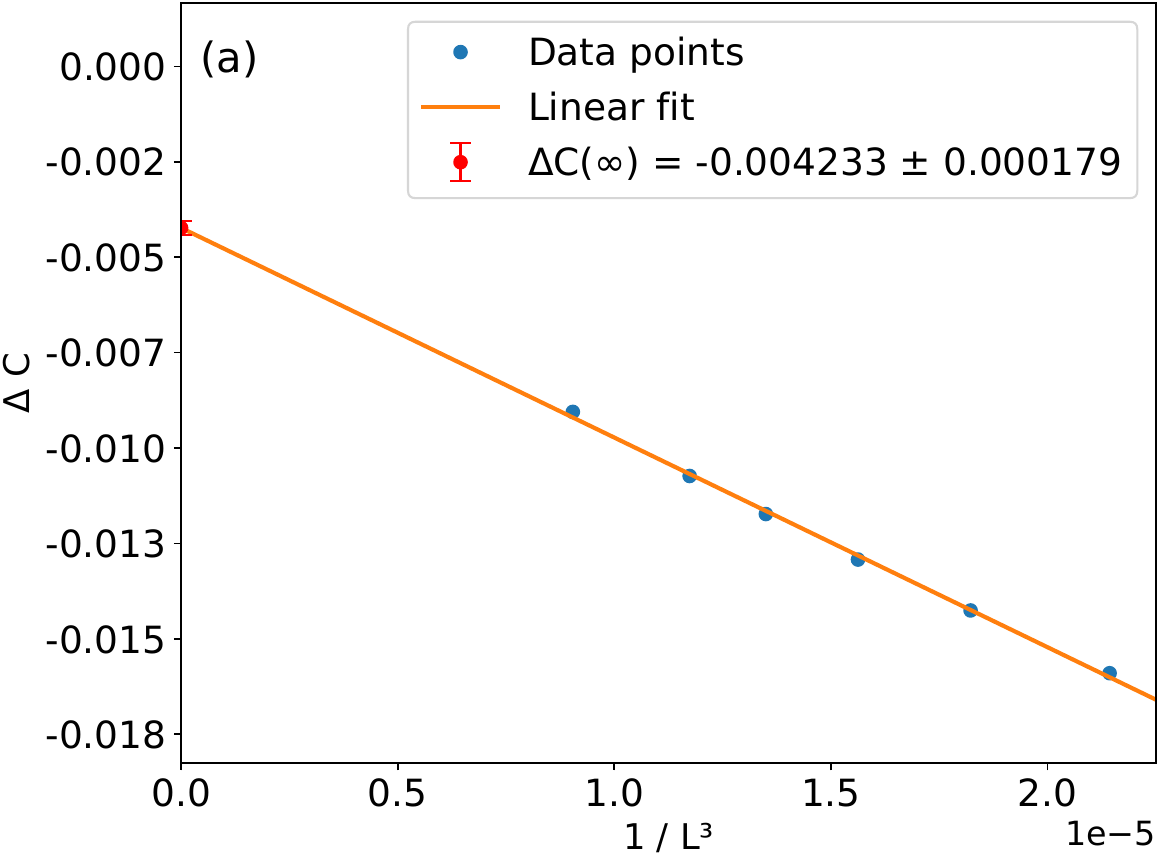}
    \end{minipage}
    \hfill
    \begin{minipage}[b]{0.32\textwidth}
        \centering
        \includegraphics[width=\linewidth]{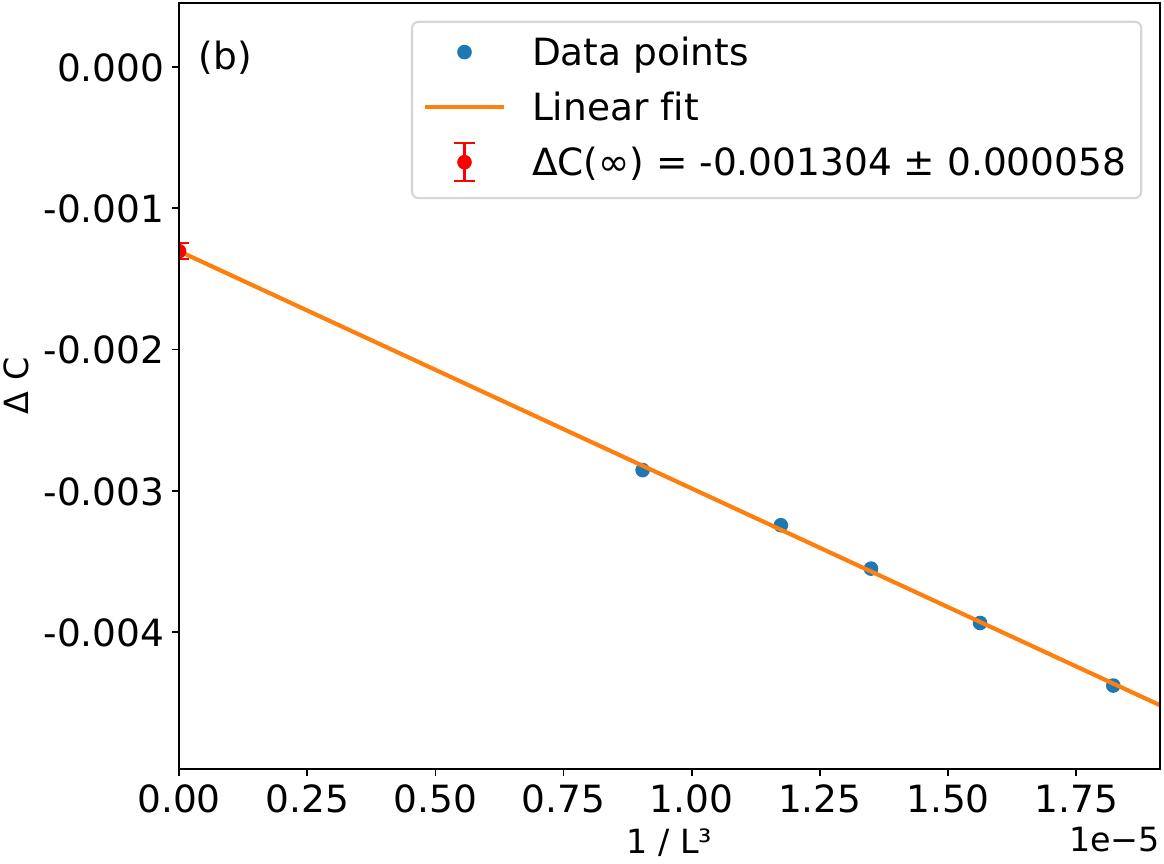}
    \end{minipage}
    \hfill
    \begin{minipage}[b]{0.32\textwidth}
        \centering
        \includegraphics[width=\linewidth]{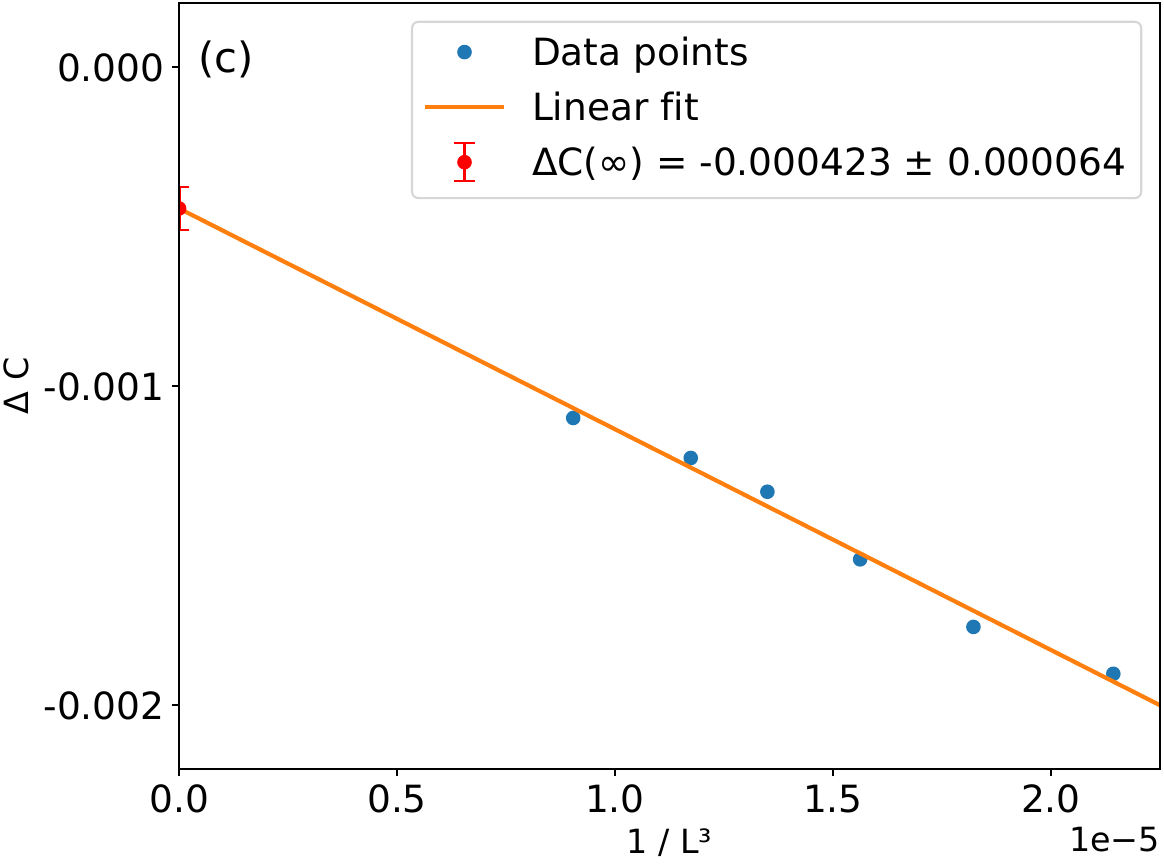}
    \end{minipage}

    \vspace{3mm} 

    \hfill
    \begin{minipage}[b]{0.32\textwidth}
        \centering
        \includegraphics[width=\linewidth]{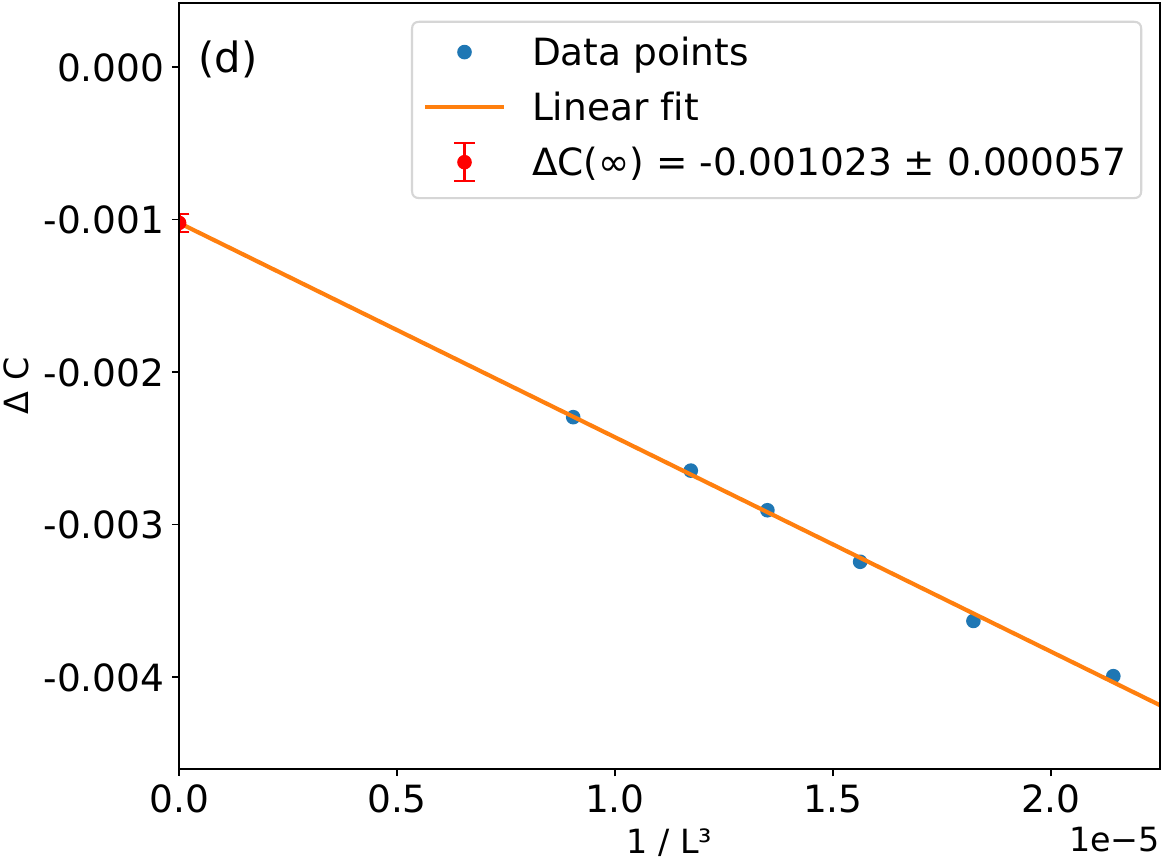}
    \end{minipage}
    \hfill
    \begin{minipage}[b]{0.32\textwidth}
        \centering
        \includegraphics[width=\linewidth]{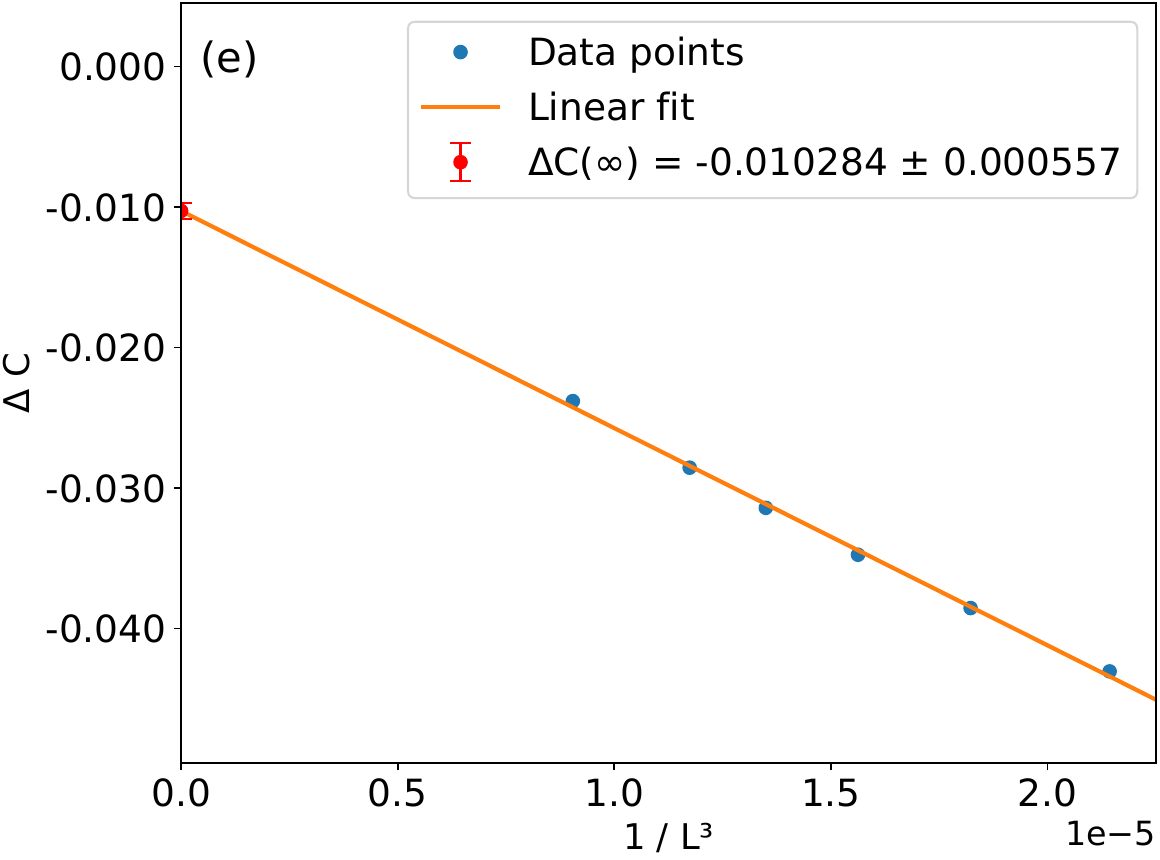}
    \end{minipage}
    \hfill
    \begin{minipage}[b]{0.32\textwidth}
        \centering
        \includegraphics[width=\linewidth]{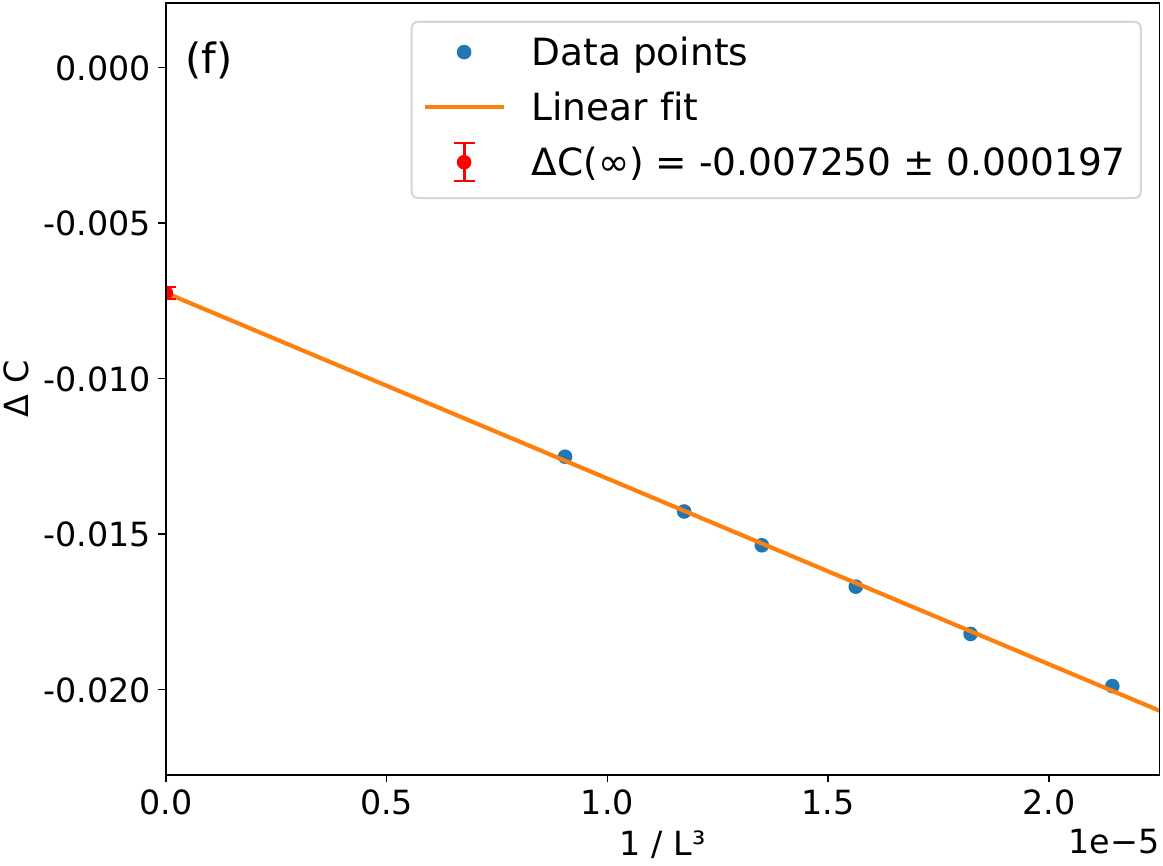}
    \end{minipage}

    \caption{Scaling of $\Delta \mathcal{C}$ with system size with the red vertical bars denoting the error in the extrapolated value. Figures (a), (b), (c) and (d) (left and central panel) correspond to impurities varying non-zero elements of the Hamiltonian matrix whereas figures (e) and (f) correspond to impurities varying zero elements of the Hamiltonian matrix (a) Onsite impurity (b)$A_0- \Gamma^1$ impurity (c) $A_0-\Gamma^2$ impurity, (d) $B_0$ impurity, (e) $S_x$ impurity, (f) Fourth nearest neighbour hopping}
    \label{fig:all_six}
\end{figure*}

\section{3D topological insulator}

To demonstrate the advantage of KPM and draw some relevance to realistic TIs, we consider a tight-binding model of time-reversal symmetric TIs in three spatial dimensions belonging to class AII. The model is obtained by regularizing the low-energy Hamiltonian of TIs like $\text{Bi}_2\text{Se}_3$ on a cubic lattice (with nearest-neighbour hopping).
The single-particle Hamiltonian in momentum space is given by \cite{Zhang09,Liu10}
\begin{align}
    H_0(\mathbf{k}) ={}& \left(M_0 + 4M_2 + 2M_1\right)\Gamma^5 \nonumber \\
    & - 2\left(M_1 \cos{k_z} + M_2(\cos{k_x} + \cos{k_y})\right)\Gamma^5 \nonumber \\
    & + B_0 \sin{k_z} \Gamma^4 
    + A_0 \left( \sin{k_y} \Gamma^1 - \sin{k_x} \Gamma^2 \right)
\label{k_space_Hamiltonian}
\end{align}
where we have adopted the following convention for gamma matrices :
\begin{equation} 
    \Gamma^1 = \sigma_x\tau_x\, , \, \Gamma^2 = \sigma_y\tau_x \, , \, \Gamma^3 = \sigma_z\tau_x \, , \, \Gamma^4 = \sigma_0\tau_y \,  , \, \Gamma^5 = \sigma_0\tau_z
\end{equation}
The corresponding lattice Hamiltonian can be obtained through a straightforward Fourier transform\cite{Chen20_absence_edge_current,Chen23_universal_marker}, whose form will be omitted for simplicity. Unless otherwise specified, we choose the parameters to be $M_1=M_2=1$, $A_0=B_0=2$ and $M_0=-2$. A topological phase transition occurs at $M_0=0$, where the bulk gap closes at $\textbf{k}= (0,0,0)$. For $M_0<0$, the system is in a topological phase and for $M_0>0$, the system is in a trivial phase. Our aim is to investigate how various sorts of impurities influence the topological order, as well as the quantum criticality near the $M_{0}=0$ critical point. The unused Dirac matrix is $W=\Gamma^3$, so the topological operator in Eq.~(\ref{topological_operator}) takes the form:
\begin{equation}
    {\hat{\mathcal C}} = -8\pi i\Gamma^3(Q\hat{x}P\hat{y}Q\hat{z}P + P\hat{x}Q\hat{y}P\hat{z}Q),
\end{equation}
where we use exponentiated position operators \cite{Oliveira24}
\begin{equation}
    \hat{x} = \frac{L_x}{2\pi}e^{\frac{2\pi i x}{L_x}} \, , \,     \hat{y} = \frac{L_y}{2\pi}e^{\frac{2\pi i y}{L_y}} \, , \,     \hat{z} = \frac{L_z}{2\pi}e^{\frac{2\pi i z}{L_z}}
\end{equation}
that have been shown to cure the anomaly of the marker around the sample edge. In homogeneous systems, the topological marker captures the strong topological invariant that describes the topological phases of the entire momentum space, in contrast to the weak topological invariants for some subspace of the Brillouin zone whose real space markers remain elusive at present.

\subsection{Disorder}

Drawing knowledge from previous works, we are particularly interested in the question of whether the spatially averaged marker remains quantized in the presence of disorder. In particular, in one- (1D) and two-dimensional (2D) TIs, it can be proved analytically based on perturbation theory that if the dilute impurities correspond to modifying nonzero matrix elements of the lattice Hamiltonian, then the average marker is conserved, indicating that the topological order is robust against this type of disorder\cite{Oliveira24}. On the contrary, if the impurities modify zero matrix elements of the Hamiltonian, then in general the average marker no longer remains a quantized integer. This issue is also of significant importance for the three-dimensional TIs addressed in the present work because of the recent experimental work on amorphous Bi$_{2}$Se$_{3}$ \cite{corbae1,Ciocys2024}. Unfortunately, at present we do not have an analytical theory to discern which types of impurities destroy the average marker in 3D class AII and which types do not, so we solely rely on the numerical results obtained through KPM to answer this question.

From Eq.~(\ref{k_space_Hamiltonian}), it is clear that there are only four types of nonzero matrix elements in the lattice Hamiltonian deriving from the corresponding momentum space Hamiltonian, namely the mass term $M_{0}$, the in-plane $A_{0}$ (which appear as coefficients of $\Gamma^1$ and $\Gamma^2$) and out-of-plane $B_{0}$ hopping terms. The corresponding impurity terms in the real space Hamiltonian can be written as:
\begin{equation}
    H^1_{\text{imp}} = M_{\text{imp}}\ket{\textbf{r}}\bra{\textbf{r}}\otimes \Gamma^5
\end{equation}
\begin{equation}
    H^2_{\text{imp}} = A_{\text{imp}}\ket{\textbf{r}}\bra{\textbf{r}+a\hat{y}}\otimes \Gamma^1 + \text{h.c}
\end{equation}
\begin{equation}
    H^3_{\text{imp}} = A_{\text{imp}}\ket{\textbf{r}}\bra{\textbf{r}+a\hat{x}}\otimes \Gamma^2 + \text{h.c}
\end{equation}
\begin{equation}
    H^4_{\text{imp}} = B_{\text{imp}}\ket{\textbf{r}}\bra{\textbf{r}+a\hat{z}}\otimes \Gamma^4 + \text{h.c}
\end{equation}
If an impurity does not correspond to any of these four types, it must modify some zero matrix element of the lattice Hamiltonian. For example, an in-plane magnetic moment can be  introduced by an onsite impurity of the form:
\begin{equation}
    H_{\text{onsite}} = M_{S_x}\ket{\textbf{r}}\bra{\textbf{r}}\otimes \sigma_x \tau_0
\end{equation}
or one can introduce an impurity term via an additional hopping in the Hamiltonian:
\begin{equation}
    H_{\text{hopping}} = t_{\text{imp}}\ket{\textbf{r}}\bra{\textbf{r}+2a\hat{y}}\otimes \Gamma^1 + \text{h.c}
\end{equation}

To illustrate the influence of the two kinds of disorder on the average marker we show the spatial pattern of the marker around a single hopping impurity in Fig.~\ref{fig:spatial_pattern_marker},  where we plot the marker in the $xy$ plane (z=24.5) that passes through the impurity site, in a cubic system of size $48^3$.  For our simulations, we have used $M_{\text{imp}}=M_{S_x}=4$, $A_{\text{imp}}=B_{\text{imp}}=3$ and $t_{\text{imp}}=1$ and a KPM expansion of order 800. From the figure, it can be seen that even though the strength of the impurity hopping  ($t_{imp}$) modifying  a zero-matrix element is less than the impurity hopping modifying a non-zero element ($A_{imp})$ of the matrix, it produces larger deviations from the average marker value of the clean system.

 We proceed to clarify whether each type of impurity modifies the quantized topological invariant in its vicinity. Since the modification to the quantized value is tiny, it can only be accurately captured by a very large lattice with KPM up to very high order, which is a difficult numerical task. Nevertheless, one can draw evidence by performing a finite size scaling by calculating the change in the total marker caused by a single impurity as a function of the inverse of the system size $1/L^{3}$
 \begin{equation}
    \Delta C = \mathcal{C}^{\text{clean}} - \mathcal{C}^{\text{imp}}.
\end{equation}
For the system with impurity, we compute the average marker in a $10 \times 10 \times 10 $ cube located at the center of the sample of size $L^3$ :
 \begin{equation}
     \mathcal{C}^{\text{imp}} = \sum_{\mathbf{r}} \mathcal{C}(\mathbf{r})
 \end{equation}
where the sum is over all the sites inside the central cube and the impurity is placed at the center of this cube. Figure \ref{fig:all_six} (a), (b), (c), and (d) show the results for the four types of impurities that modify nonzero matrix elements $(M_{0},A_{0}-\Gamma^1,A_{0}-\Gamma^2,B_{0})$, which show numerical evidence that the modification of the marker $\Delta{\cal C}$ extrapolates to less than $0.001$ in the thermodynamic limit $1/L^{3}\rightarrow 0$, suggesting that they do not change the global topological order just like what happens in 1D and 2D TIs. In contrast, for the other two types of impurities that modify zero matrix elements shown in Fig.~\ref{fig:all_six} (e) and (f), the modification to the average marker saturates to a small but nonzero value at $1/L^{3}\rightarrow 0$, of the order of $0.01$, suggesting that the topological order is altered by the presence of these types of impurities. We conclude that the response of 3D TIs to a single impurity is similar to 1D and 2D TIs, i.e., the impurity that modifies the zero matrix elements of the Hamiltonian changes the topological invariant, whereas the impurity that modifies the nonzero matrix elements does not, suggesting a universal response to impurities across different dimensions.

\begin{figure}[htbp]
    \centering
    \includegraphics[width=0.9\linewidth]{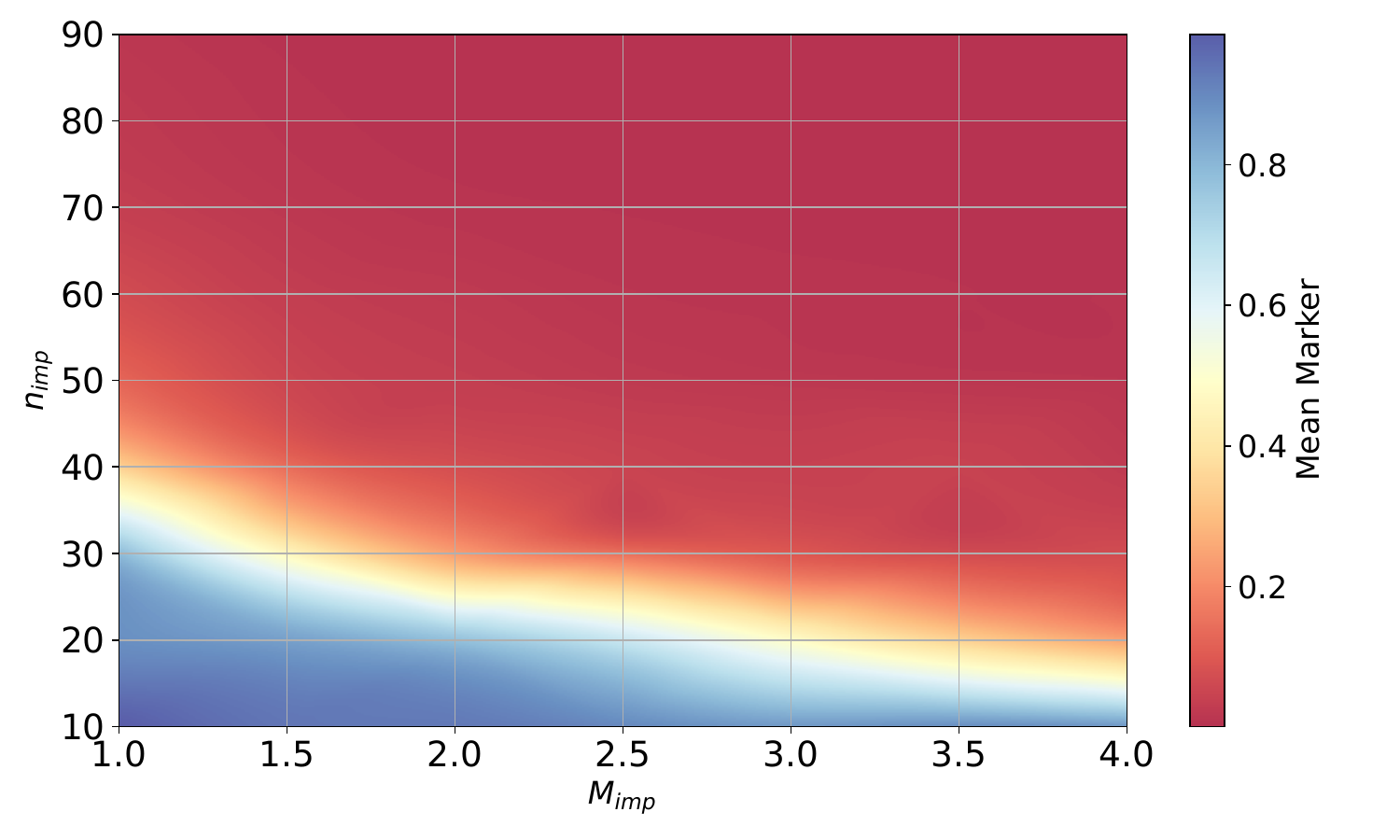}
    \label{fig:fphase_diag_fig1}
    \vspace{0.3cm} 
    \includegraphics[width=0.9\linewidth]{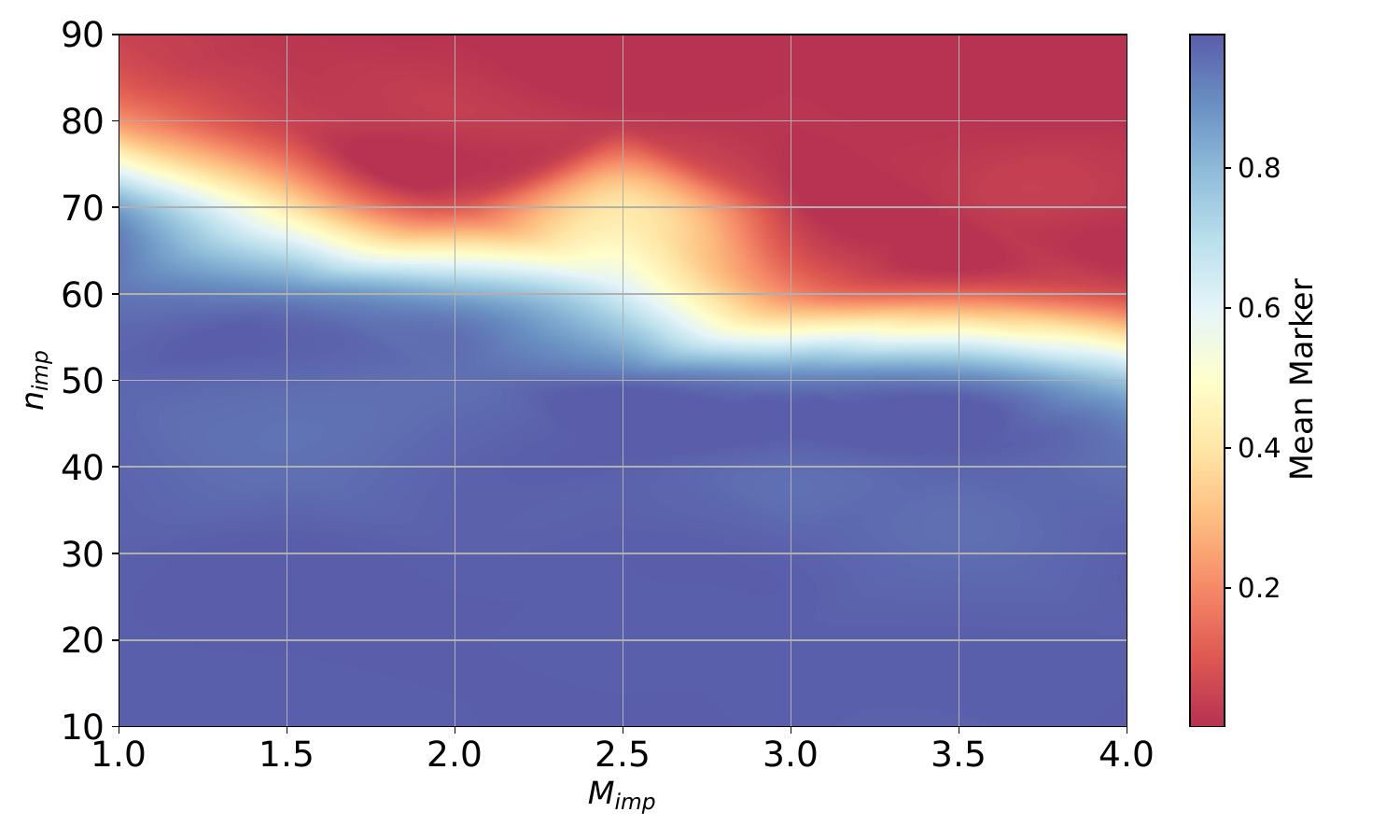}
    \caption{Phase diagram as a function of concentration of impurities for different mass parameters. Top panel : $M_0=-0.5$. Bottom panel : $M_0=-2.0$}
    \label{fig:phase_diag_fig2}
\end{figure}

 We proceed to examine a feature that has been reported for 1D and 2D TIs, namely the disorder-induced smooth crossover between two topological phases. Intuitively, because the value of the mass term $M_{0}>0$ or $M_{0}<0$ controls the topological order, if the mass term randomly changes sign on some lattice sites, the system's behaviour will be similar to  that of an alloy of two topological phases and therefore the spatially averaged topological marker can take some fractional value. On the other hand, in Fig.~\ref{fig:all_six} we have already shown that mass term impurities modify nonzero matrix elements of the Hamiltonian and hence the average marker should remain unchanged. Consequently, one expects that these mass term impurities can change the topological phases only if their density $n_{imp}$ or the impurity strength $M$ is high enough. In Ref.~\onlinecite{Oliveira24}, it is proposed that the relevant length scale for the impurity density is the correlation length $\xi\sim A_{0}/M_{0}$ determined by the Fermi velocity and bulk gap, and the relevant energy scale for the impurity strength is the bulk gap itself $M_{0}$. In Fig.~\ref{fig:phase_diag_fig2}, we present the phase diagram in the presence of the mass term impurities, where we plot the average marker as a function of impurity density and strength. Comparing small $M_{0}=-0.5$ and large $M_{0}=-2$ bulk gap, we see that later requires a higher impurity density and strength to cross into the topologically trivial phase. This result agrees with the empirical formula that the topological order is destroyed only if the impurity density is higher than that set by the correlation length or if the impurity strength is larger than the bulk gap
 \begin{eqnarray}
     \frac{L^{3}}{n_{imp}N}\lesssim\xi^3,\;\;\;| M_{imp}| \gtrsim |M_{0}|,
     \label{xi_M0_empirical}
 \end{eqnarray}
similar to what occurs in 1D and 2D TIs\cite{Oliveira24}.

\begin{figure}[htbp]
    \centering    
    \includegraphics[width=0.9\linewidth]{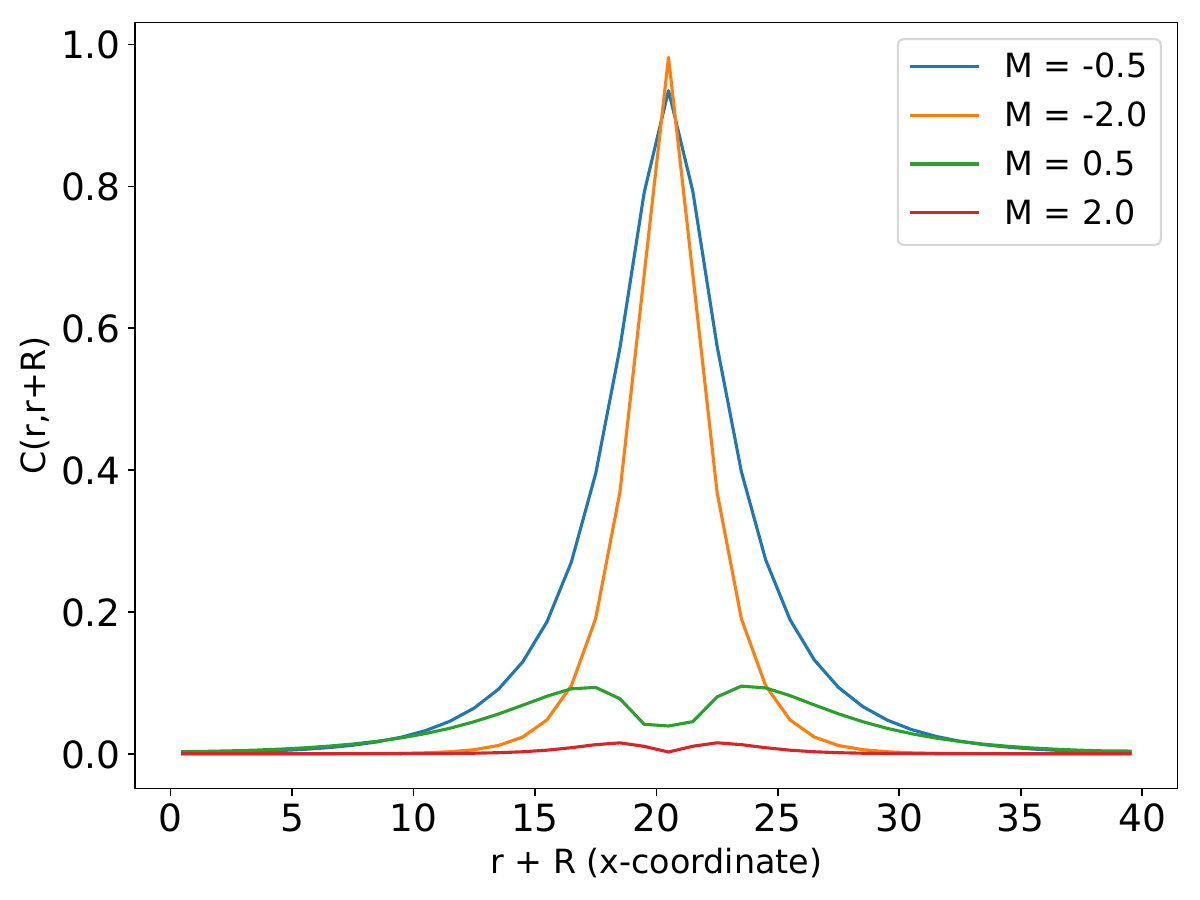}
    \caption{Non-local topological marker for different values of mass parameter $M$}
    \label{fig:pattern_non_local_marker}
\end{figure}

Finally, we present the numerical result of the nonlocal topological marker ${\cal C}({\bf r,r+R})$ in homogeneous 3D TIs as a function of the separation ${\bf R}$ in Fig.~\ref{fig:pattern_non_local_marker}. The significance of this nonlocal marker is that it quantifies the quantum criticality near topological phase transitions, owing to the fact that it is equivalent to the Fourier transform of the curvature function that integrates to the topological invariant in momentum space\cite{Molignini23_Chern_marker,Chen23_universal_marker}. Because the curvature function narrows and diverges in momentum space as the system approaches a topological phase transition, the nonlocal marker becomes more and more long ranged accordingly. The results in Fig.~\ref{fig:pattern_non_local_marker}  confirm this Fourier transform picture, where one sees that the decay length of ${\cal C}({\bf r,r+R})$ as a function of ${\bf R}$ gradually increases as the mass terms gradually approach the critical point $M_{0}\rightarrow 0$. This is because the decay length of ${\cal C}({\bf r,r+R})$ is simply given by the correlation length\cite{Molignini23_Chern_marker,Chen23_universal_marker} $\xi$ in Eq.~(\ref{xi_M0_empirical}), which diverges at the critical point\cite{Chen17,Chen19_universality} $\lim_{M_{0}\rightarrow 0}\xi=\infty$ with a critical exponent $\nu=1$ in homogeneous systems. As a result, the decaying profile of the nonlocal marker near the critical point can only be correctly captured if the system size is larger than the correlation length $L\gtrsim\xi$, which would not be possible to investigate numerically without employing KPM. Thus our analysis indicates that the KPM also helps to quantify the topological quantum criticality in higher dimensions, especially that it enables the investigation of the critical behavior much closer to the critical point. Finally, we remark that it is also possible to calculate the nonlocal marker for disordered systems and extract the critical exponent, an issue that has been raised recently within the context of a Chern-Chern correlation function\cite{Favata25}. This issue remains to be further explored.

\section{Conclusions}

In summary, we elaborate the numerical efficiency of the KPM in calculating the universal topological marker and the nonlocal marker, rendering a powerful tool that can accurately capture the disorder effect and topological quantum criticality in lattice models of TIs and TSCs in higher dimensions. To draw relevance to prototype TIs like Bi$_{2}$Se$_{3}$ and Bi$_{2}$Te$_{3}$, we apply this method to 3D class AII, and reveal that impurities that correspond to altering nonzero matrix elements of the lattice Hamiltonian do not change the spatially averaged marker, regardless of whether the impurities break the time-reversal symmetry or not. In contrast, impurities that alter the zero matrix elements of the lattice Hamiltonian deteriorate the average marker. Our work thus clarifies the robustness of bulk topological order in real TIs like Bi$_{2}$Se$_{3}$ and Bi$_{2}$Te$_{3}$ against different types of impurities. Furthermore, impurities that locally change the mass term can be used to continuously tune the average marker from one integer to another, rendering a smooth crossover between topological phases. All these features are very similar to those in lower dimensional TIs and TSCs, suggesting that different topological materials may share a common response to disorder. Finally, the KPM also captures the increasing decay length of the nonlocal topological marker as the system approaches the topological phase transitions, and hence helps to quantify the topological quantum criticality in higher dimensions. We anticipate that this KPM plus universal topological marker technique to other symmetry classes in 3D and to higher dimensions can help address whether these disorder effects are truly universal, which remains to be further explored.

\acknowledgments{R.R. is supported by Center for Emergent Materials at the Ohio State University, a National Science Foundation (NSF) MRSEC through NSF Award No. NSF DMR-2011876. W.C. acknowledges the support of the INCT project Advanced Quantum Materials,
involving the Brazilian agencies CNPq (Proc. 408766/2024-7), FAPESP, and CAPES, as well as the financial support of the productivity in research fellowship from CNPq.}

\section*{Data Availability}
The data that support the findings of this article are openly available \cite{data}.

\end{document}